\documentstyle[aps,prl,multicol,epsfig]{revtex}

\begin{document}

\title{Electro-magnetic Aharonov-Bohm effect in a 2-D electron gas ring}

\author{W.G. van der Wiel$^{1,*}$, Yu.V. Nazarov$^{1}$, S. De Franceschi$^{1}$, T.
Fujisawa$^{2}$,\\
J.M. Elzerman$^{1,*}$, E.W.G.M. Huizeling$^{1}$, S.
Tarucha$^{2,3,*}$ and L.P. Kouwenhoven$^{1,*}$}
\address{$^1$Department of Applied Physics and DIMES, Delft University of Technology, PO Box 5046,
2600 GA Delft, The Netherlands}
\address{$^2$NTT Basic Research Laboratories, Atsugi-shi, Kanagawa 243-0198, Japan}
\address{$^3$Department of physics, University of Tokyo, 7-3-1, Hongo, Bunkyo-ku, Tokyo 113-0033, Japan}
\address{$^*$also at ERATO Mesoscopic Correlation Project}

\maketitle

\begin{abstract}
We define a mesoscopic ring in a 2-dimensional electron gas (2DEG)
interrupted by two tunnel barriers, enabling us to apply a
well-defined potential difference between the two halves of the
ring. The electron interference in the ring is modified using a
perpendicular magnetic field and a bias voltage. We observe clear
Aharonov-Bohm oscillations up to the quantum Hall regime as a
function of both parameters. The electron travel time between the
barriers is found to increase with the applied magnetic field.
Introducing a scattering model, we develop a new method to measure
the non-equilibrium electron dephasing time, which becomes very
short at high voltages and magnetic fields. The relevance of
electron-electron interactions is discussed.
\end{abstract}


\begin{multicols}{2}
%\section{Introduction}
%\label{sec:intro}

Aharonov and Bohm predicted that the phase of an electron wave is
affected by both magnetic and electric potentials, being
observable in an interference experiment \cite{AB59}. A magnetic
flux, $\Phi$, threading a loop leads to an oscillating
contribution to the conductance with period $\Phi _{0} = h/e$: the
{\it magnetic} Aharonov-Bohm (AB) effect. In a solid state device,
AB oscillations are observable at mesoscopic scale, where quantum
coherence is preserved. The effect was first observed in a metal
ring \cite{Webb85} and later in 2DEG rings \cite{Timp87}.

In the case of the {\it electrostatic} AB effect, as originally
suggested, the electron phase is affected by an electrostatic
potential, although the electrons do not experience an electric
field \cite{AB59}. The required geometry is difficult to realize
and so far only geometries have been investigated where an
electric field changes the interference pattern
\cite{Petrashov,Vegvar,Krafft}. In Ref. \cite{Nazarov93} a ring
geometry interrupted by tunnel barriers was suggested, where a
bias voltage, $V$, leads to an electrostatically controlled AB
effect with predictions very similar to the original proposal
\cite{AB59}. This effect was observed in a disordered metal ring
with two tunnel junctions \cite{Oudenaarden98}.

In this Letter, we report the observation of an electrostatic AB
effect in a quasi-ballistic ring-shaped 2DEG system interrupted by
tunnel barriers. There are two important distinctions from metal
systems that play a key role in the present study. First, only a
few electron modes are involved in transport, in contrast to tens
of thousands in a typical metal wire. Second, in a 2DEG we can
enter the quantum Hall regime using a perpendicular magnetic
field, $B$.

We observe both a magnetic and electrostatic AB effect up to the
edge channel regime (filling factor $\nu$ = 3). The electron
travel time between the barriers increases with $B$. We ascribe
this to the bending effect of the Lorentz force on the electron
trajectories. Importantly, the electrostatic oscillations form a
new tool to determine the non-equilibrium electron dephasing time,
$\tau_{\phi}$. The possibility to estimate the electron dephasing
time at a controlled, finite energy and magnetic field
distinguishes this method from weak localization experiments used
to estimate the (equilibrium) electron dephasing time
\cite{Mohanty97}.
%fig1
\begin{figure}[htbp]
  \begin{center}
  \centerline{\epsfig{file=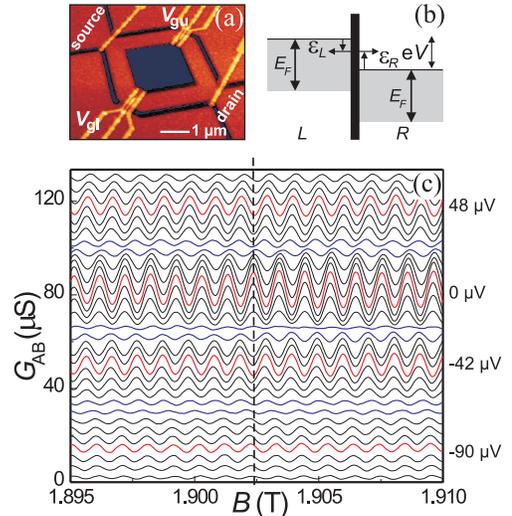, width=6.5cm, clip=true}}
    \caption{(a) Atomic force microscope image of the device. The AB ring
    is defined in a 2DEG by dry etching (dark regions, depth $\sim$ 75 nm). The
    2DEG with electron density $3.4\times 10^{15} m^{-2}$ is situated 100 nm below the
    surface of an AlGaAs/GaAs heterostructure. In both arms of the ring
    (lithographic width 0.5 $\mu$m; perimeter 6.6 $\mu$m) a barrier can be defined
    by applying negative voltages to the gate electrodes $V_{gu}$ and $V_{gl}$. The
    other gates are not used. (b)Schematic energy picture at one of the barriers (see text).
    (c) Differential AB conductance, $G_{AB}$, as function of $B$ at 15 mK for different
    source drain voltages, $V$, separated by 5.3$\mu$V. The traces have a vertical offset.
    Three values for $V$ are indicated on the right. The dashed line highlights a phase
    change by a change in the amplitude and sign of the AB oscillations,
    indicating an electrostatic AB effect.}
  \end{center}
  \label{fig1}
\end{figure}
\noindent We find that $\tau_{\phi}$ decreases as a function of
$B$ and $V$ and reaches $\tau_{\phi} |\varepsilon| \simeq \hbar$
at the highest $B$, with $\varepsilon$ the electron energy
measured from the Fermi level. This demonstrates the increasing
importance of many-body effects at quantizing magnetic fields
where the Fermi-liquid picture and the quasi-particle concept are
at the edge of applicability.

Our device (Fig. 1a) consists of an AB ring defined in a 2DEG
\cite{Wiel00}. The gate electrodes $V_{gu}$ and $V_{gl}$ define a
barrier in each arm of the ring. The effective width of the arms
supports $N_{tr} \simeq$ 10 transport channels (i.e. the
conductance of the ring without barriers is approximately
$10e^{2}/h$ at $B$ = 0 T). In addition to the dc bias voltage,
$V$, we apply a relatively small ac voltage (3 to 10 $\mu$V)
between source and drain contacts. We measure the differential
conductance, $G=dI/dV$, using a lock-in technique in a dilution
refrigerator with a base temperature of 15 mK.

Figure 1c shows the differential AB conductance, $G_{AB}$, versus
$B$ and $V$, around 1.9 T. $G_{AB}$ is extracted from the measured
$G$, by fitting a polynomial to the smoothly varying background in
$G$. Subtraction of this polynomial yields the oscillating part,
$G_{AB}$. The barriers are tuned such that the conductance through
one arm is 0.2 $e^{2}/h$. $G_{AB}$ as function of magnetic field
exhibits AB oscillations with a period $B_{0}$ = 1.0 mT, in good
agreement with $h/(eS)$, where $S$ is the area enclosed by the
ring. The effect of the dc bias voltage, $V$, is schematically
shown in Fig. 1b. The transmitted part of the electron wave has
%fig2
\begin{figure}[htbp]
  \begin{center}
  \centerline{\epsfig{file=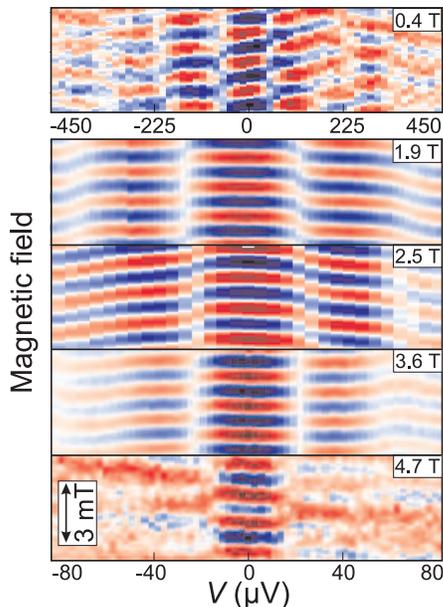, width=5.8cm, clip=true}}
    \caption{Color scale plots of the AB conductance in the plane of
    magnetic field (vertical axis) around different $B$ values and voltage, $V$
    (horizontal axis). Red (blue) corresponds with maximum (minimum)
    AB conductance. White corresponds with zero AB conductance.
    The phase-changes indicate the electrostatic period which
    varies from $V_{0}$ = 225 $\mu$V at $B$ = 0.4 T to $V_{0}$ = 60 $\mu$V at $B$ = 4.7 T.}
    \label{fig2}
  \end{center}
\end{figure}
\noindent energy $\varepsilon_{R}$, whereas the reflected part has
$\varepsilon_{L}$ with respect to the Fermi level in each half of
the ring (note $eV$ = $\varepsilon_{R}$ - $\varepsilon_{L}$ and
$|\varepsilon_{L}|$,$|\varepsilon_{R}|$,$|eV| \ll E_{F}$). The
energy difference leads to a phase difference $\Delta \Phi _{V}$
$=2\pi eVt_{0}/h$, where $t_{0}$ is the ($B$-dependent) time the
reflected and transmitted electron waves spend in their respective
parts of the ring before they interfere \cite{Nazarov93}. We thus
expect electrostatic AB oscillations by changing $V$.

The dashed line in Fig. 1c highlights the effect of the applied dc
voltage, $V$. The red curve at $V$ = 48 $\mu$V has a minimum when
it crosses the dashed line. By decreasing $V$ the amplitude of the
AB oscillations decreases. The two blue curves near $V$ = 24
$\mu$V show a case where the amplitudes are small and the sign of
the amplitude changes. At $V$ = 0 $\mu$V, the amplitude is
maximal. Comparing the 48 and 0 $\mu$V traces, the AB oscillation
has acquired a change in phase by $\pi$. This continues by another
phase shift of $\pi$ when $V$ is decreased to -42 $\mu$V. Thus,
the electrostatic period, $V_{0}$, is 90 $\mu$V at this magnetic
field. This corresponds to an electron travel time $t_{0}$ = 45
ps. We note that the Onsager-B\"{u}ttiker symmetry relation $G(B)
= G(-B)$ for a 2-terminal measurement \cite{Buttiker88} only holds
at small bias voltages. Therefore, the phase of the oscillations
at the field of the dashed line is not restricted to 0 or $\pi$.

Figure 2 shows a color scale plot of $G_{AB}$ versus $B$ and $V$
for five different $B$ ranges. The main features of the results
are as follows. The amplitude of the AB oscillations is relatively
small ranging from $\sim$0.01$G$ to $\sim$0.1$G$. The period of
the oscillations in $G_{AB}(V)$ decreases with $B$, as is clearly
seen by comparing the different panels in Fig. 2. For $B$ = 0.4 T,
1.9 T, 3.6 T, and, less clearly, for $B$ = 4.7 T, the phase of the
magnetic oscillations exhibits sharp $\pi$-shifts at lower
voltages. At higher voltages, the phase changes more smoothly. At
$B$ = 2.5 T, the phase changes smoothly over the entire voltage
range. We return to the phase evolution at different $B$ and $V$
later, when we discuss our scattering model.

The AB amplitude decreases with $V$ and, the higher $B$, the
faster the decrease. The decreasing AB amplitude is clearly seen
in Fig. 3 where we plot the normalized AB conductance,
$G_{AB}(V)/G_{AB}(V=0)$, versus $V$ at $B$ = 1.9 T. We stress that
Fig. 3 is obtained by taking a single horizontal cut, i.e. for a
single $B$ with no averaging \cite{average}, from the $B$ = 1.9 T
panel of Fig. 2 (but over a larger voltage range). Note that
$G_{AB}(V)/G_{AB}(V=0)$ can be negative since it only represents
the oscillating part of the total conductance. The curve is
approximately symmetric in $V$ ($G_{AB}(V) \approx G_{AB}(-V)$),
indicating that within the range of applied voltages the electron
travel time, $t_0$, hardly changes with $V$. Thus, the
electrostatic period does not vary within the applied bias window
and we cannot attribute the decrease with $V$ in Fig. 3 to
self-averaging of trajectories with different periods. We
therefore believe that the only possible mechanism for the
decrease is electron dephasing due to inelastic processes.

%Theoretical model
The small relative amplitude of the AB oscillations indicates
strong elastic scattering in the arms, whereas the quickly
decreasing AB amplitude with $V$ indicates strong inelastic
scattering. Since these scattering mechanisms were not included in
the model of Ref. \onlinecite{Nazarov93}, we introduce a simple
interference model including dephasing, proceeding along the lines
of the general scattering formalism \cite{Buttiker86}.

We assume single-channel tunnel junctions with transmission
amplitudes $t_{1}$ and $t_{2}$, respectively. A small AB amplitude
implies that the amplitudes of propagating waves between the
junctions, $r_{12}^{L}$ or $r_{21}^{R}$ (see Fig. 4), are small so
that we only consider scattering processes of first order in $r$.
The oscillating part of the transmission probability, $T_{AB}$, is
built up from the pairwise interference of the electron
trajectories shown in Fig. 4. Collecting the contributions at a
given energy, we obtain
\begin{equation}
T_{AB} = 2 {\rm Re}(t_1t^*_2 e^{i\Phi_{AB}} (r^L_{12} + r^{*L}_{21}) ( r^R_{21} + r^{*R}_{12}))
\end{equation}
We assume right-left symmetry \cite{RLsymmetry} and specify the
energy dependence of the propagation amplitudes as follows
\begin{equation}
r^L_{12(21)} = r_{- (+)}e^{i\varepsilon_L t_0/\hbar}; \
r^R_{12(21)} = r_{+ (-)}e^{i\varepsilon_R t_0/\hbar}
\end{equation}
where $r_{+(-)}$ correspond to (counter)clockwise propagation of
electrons along the ring (remind that $eV$ = $\varepsilon_{R}$ -
$\varepsilon_{L}$). For the differential conductance this yields
(assuming $t_1=t_2$)
\begin{eqnarray}
G_{AB}(V)/G= |r_{-}|^2 \cos(eVt_0/\hbar) \cos(2 \pi \Phi/\Phi_{0}+\Phi_{a})+ \nonumber \\
|r_{+}|^2 \cos(eVt_0/\hbar) \cos(2 \pi \Phi/\Phi_{0} + \Phi_{b}) + \nonumber \\
2 |r_{+}r_{-}|[\cos(eVt_0/\hbar + \Phi_c) + \nonumber \\
\sin(eVt_0/\hbar+\Phi_c)eVt_0/\hbar] \cos(2 \pi \Phi/\Phi_{0} +
\Phi_{d}) \label{general_formula}
\end{eqnarray}
where $\Phi_{a,b,c,d}$ are contributions to the phase that vary
slowly with $B$ in comparison to $\Phi/\Phi_{0}$ (e.g. due to
bending of the electron trajectories). At small $B$ time
reversability holds, implying $r_{+}=r_{-}$ and $\Phi_{a,b,d} =0$.
Under these conditions, the phase of the magnetic oscillations
only changes by $\pi$ as a function of $V$. Interestingly, the
same happens at high $B$. In this case Lorentz bending suppresses
counterclockwise propagation, so that $r_{+}\gg r_{-}$. The result
is then dominated by the second term in Eq. \ref{general_formula}.
For a $B$-interval in which $\Phi_{b}$ can be considered constant,
we find again that the phase of the magnetic oscillations only
changes by $\pi$ as a function of $V$. In the intermediate regime
$r_{+}\simeq r_{-}$, and $\Phi_{a,b,c,d}$ vary slowly with $B$.
The phase of the magnetic oscillations in this case continuously
shifts with voltage, saturating at  $ V \gg \hbar / t_0 e$.

To model electron dephasing due to inelastic processes, we assume
that the electron amplitudes at a given energy are suppressed by a
factor exp$(-t/2\tau _{\varphi })$,
 $\tau _{\varphi }(\varepsilon)$ being the energy-dependent dephasing
time. In analogy with results found for a disordered electron gas
in the quantum Hall regime \cite{Polyakov98}, disordered
metal-like systems \cite{Altshuler85}, composite fermions
\cite{Lee96} and a Luttinger liquid \cite{GlazmanNATO}, we propose
a dephasing time proportional to energy, $\hbar /\tau _{\varphi
}(\varepsilon)=2\alpha \varepsilon$. Here $\alpha$ is a
dimensionless factor of the order of the dimensionless 2DEG
conductance, $G/(e^2/h)$. A small value for $\alpha$ ($\alpha \ll
1$) and large conductances ($G\gg e^{2}/h$) correspond to
vanishing electron-electron interactions \cite{Altshuler85}. On
the contrary, $\alpha \simeq 1$ and $G \simeq e^{2}/h$ signal the
importance of many-body effects. In the limit of $r_{+}\gg r_{-}$
our model for the dephasing gives
\begin{eqnarray}
G_{AB}(V)/G = |r_{+}|^2 [\cos(eVt_0/\hbar)- \nonumber \\
\alpha \sin(e|V|t_0/\hbar)] e^{-\alpha e|V|/\hbar}
 \cos(2 \pi \Phi/\Phi_{0} + \Phi_{b})
\label{fit_formula}
\end{eqnarray}

%\section{Discussions}
Our scattering model provides an explanation for the observed
experimental results. The model accounts for the abrupt
$\pi$-phase changes observed in Fig. 2 at the lowest ($B$ = 0.4 T
and 1.9 T) and highest ($B$ = 3.6 T and 4.7 T) magnetic fields
(provided $V$ is small). In the intermediate regime ($B$ = 2.5 T)
the model explains also why the phase varies more continuously.
The well-defined period $t_0$ probably indicates the formation of
an edge channel connecting the junctions. However, the magnitude
of the AB oscillations is small. This signals a strong scattering
to and from the edge channel involving almost localized states at
higher magnetic field. One can estimate $|r_{+}|^2$ as a classical
probability, assuming uniform distribution
%fig3
\begin{figure}[htbp]
  \begin{center}
    \centerline{\epsfig{file=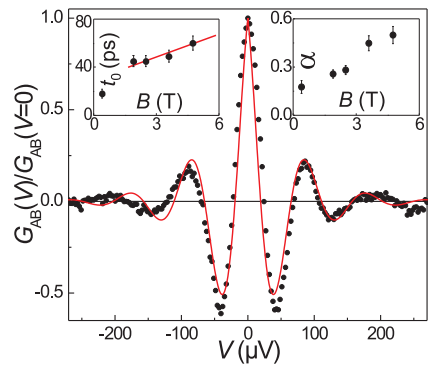, width=6.5cm, clip=true}}
    \caption{Normalized AB conductance, $G_{AB}(V)/G_{AB}(V=0)$, versus
    $V$ at $B$ = 1.9 T.The amplitude decreases rapidly with $V$ and,
    contrary to the magnetic AB effect, only a few oscillations can
    be observed. The red curve is a fit with Eq. 4, $\alpha = 0.26$,
    $t_0 = 45$ ps. The left inset shows the values of the travel time
    $t_0$ versus $B$ as extracted from fits with Eq. 4, varying from
    18 ps at $B$ = 0.4 T to 60 ps at $B$ = 4.7 T ($L/v_F$ = 13 ps,
    with $L$ the distance between the barriers and $v_F$ the Fermi
    velocity). The values of $t_0$ in the edge channel regime are
    fitted to a line with a slope of 6 ps/T. The right inset shows the
    extracted values for $\alpha = \hbar/2\varepsilon\tau_{\phi}(\varepsilon)$.
    The values of $\alpha$ increase with $B$, indicating an increasing
    importance of many-body effects at quantizing magnetic fields.}
    \label{fig3}
  \end{center}
\end{figure}
\noindent over the transport channels in the ring and $1/3$
suppression due to scattering near the openings to the source and
drain leads. This gives $|r_{+}|^2 \simeq 1/3 N_{tr} = 0.03$,
which is in agreement with the experimental value of
$G_{AB}(V=0)/G \simeq 0.02$ at low $B$.

We use Eq. \ref{fit_formula} to fit the experimental
$V$-dependence of the AB amplitude, as shown in Fig. 3 for the
particular case $B$ = 1.9 T. The reasonable quality of the fit
supports our model for the dephasing time. From such fits we
extract values for $t_0$ and $\alpha$ at various magnetic fields.
The left inset to Fig. 3 shows our results for the $B$-dependence
of $t_0$. The values in the edge channel regime are fitted to a
straight line with a slope of 6 ps/T. The increase of the travel
time can be attributed to the decreasing drift velocity,
$v_{drift}=|{\bf E}| /B$, $|{\bf E}|$ being the modulus of the
confining electric field at the arm edges. An estimate based on
parabolic confinement gives $|{\bf E}| \simeq 1-4$ $10^{5}$ V/m.
This implies a slope $d t_0/dB$ between 5 and 20 ps/T, which is in
reasonable agreement with the observed value.

In the right inset to Fig. 3 we show our results for $\alpha =
\hbar/2\varepsilon\tau_{\phi}(\varepsilon)$. The $\alpha$ values
are rather high and increase with $B$. At $\varepsilon$ = 10
$\mu$eV, we find $\tau_{\phi}$ = 180 ps at $B$ = 0.4 T and
$\tau_{\phi}$ = 65 ps at $B$ = 4.7 T. The usual Fermi-liquid
theory assumes well-defined quasi-particles, corresponding to
$\alpha \ll 1$. The fact that we observe $\alpha \simeq 1$ signals
the importance of many-body effects (electron-electron
interactions) in our sample. The Fermi-liquid theory here is on
the edge of applicability. We attribute this to significant
scattering in the arms of the ring. For a clean 2DEG one expects
no significant many-body effects until $\nu < 1$ ($B_{\nu = 1}$ =
14 T in our 2DEG). One can consider $\alpha$ as an effective
dissipative conductance in units
%fig4
\begin{figure}[htbp]
  \begin{center}
    \centerline{\epsfig{file=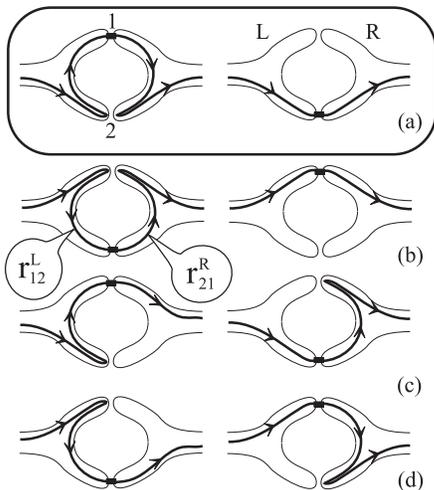, width=5.8cm, clip=true}}
    \caption{Schematic diagrams of interfering electron trajectories contributing (in first order) to the
    oscillating part of the transmission probability. At $B = 0$, the pairs of trajectories shown
    in (a)-(d) contribute equally, whereas at high magnetic field only the pair shown in (a) is relevant
    ({$\it \bf B$} points downwards). }
    \label{fig4}
  \end{center}
\end{figure}
\noindent  of $e^{2}/h$ \cite{Polyakov98,Altshuler85}. In our
case, the relevant conductance is that of the arms of the AB ring,
which is of the order of $e^{2}/h$ at the highest magnetic fields.
This is consistent with the observed $\alpha$ values.

In conclusion, using an electro-magnetic AB effect, we find a new
method to determine the non-equilibrium electron dephasing time in
a 2DEG, which becomes very short at high voltages and magnetic
field.

%\acknowledgements
We thank G. Seelig, M. B\"{u}ttiker, T. Hayashi, A. van
Oudenaarden and R. Schouten for their help. We acknowledge
financial support from the Specially Promoted Research
Grant-in-Aid for Scientific Research; the Ministry of Education,
Science and Culture in Japan; the Dutch Organization for
Fundamental Research on Matter; the New Energy and Industrial
Technology Development Organization Joint Research Program
(NTDP-98); and the European Union through a Training and Mobility
of Researchers Program network.

\end{multicols}

\begin{references}
\bibitem{average} In Ref. \cite{Oudenaarden98} the correlation function $\langle
G_{AB}(B,V)G_{AB}(B+\Delta B,V+\Delta V)\rangle$ is calculated,
where the angle brackets denote an ensemble average. For a given
$\Delta V$, one averages over an ensemble of pairs of $G_{AB}-B$
traces, each pair having a $\Delta V$ voltage difference. This
recipe causes that, if the potential landscape fluctuates on the
scale of $\Delta V$, and hence affects the (electrostatic) AB
period, the correlation function decreases.
\bibitem{RLsymmetry} This is a reasonable assumption, since in
our experiment we tune the barriers to have equal transmission.
\end{references}
\end{document}